\documentclass[twocolumn,amsmath,amssymb]{revtex4}
\usepackage[T1]{fontenc} 
\usepackage[utf8]{inputenc}
\usepackage{graphicx,multirow,subfigure}
\usepackage{float}
\usepackage{bm}
\usepackage{hyperref}
\usepackage{amsmath}
\usepackage{amssymb}
\usepackage{amscd}
\usepackage{latexsym}
\usepackage{slashed}
\usepackage{color}
\usepackage{graphicx}
\usepackage[normalem]{ulem}
\usepackage{color}
\definecolor{orange}{cmyk}{0,0.5,1,0}
\usepackage{verbatim}
\usepackage{rotating}
\newcommand{\mat}[1]{\begin{pmatrix} #1 \end{pmatrix}}

\def\lsim{\raise0.3ex\hbox{$\;<$\kern-0.75em\raise-1.1ex\hbox{$\sim\;$}}}
\def\gsim{\raise0.3ex\hbox{$\;>$\kern-0.75em\raise-1.1ex\hbox{$\sim\;$}}}
\def\met{\slashed E_T}
\usepackage{cancel}
\usepackage[utf8]{inputenc}

\def\be{\begin{equation}}
\def\ee{\end{equation}}
\def\bea{\begin{eqnarray}}
\def\eea{\end{eqnarray}}

\begin{document} 
\title{Long-Lived BLSSM Particles at the LHC}
\author{W. Abdallah$^{1,2}$, A. Hammad$^{3}$, A. Kasem$^{4,5}$ and S. Khalil$^{6}$}
\vspace*{0.2cm}
\affiliation{$^1$Department of Mathematics, Faculty of Science, Cairo University, Giza 12613, Egypt.\\
$^2$Harish-Chandra Research Institute, Chhatnag Road, Jhunsi, Allahabad 211019, India.\\
$^3$Department of Physics, University of Basel, Klingelbergstra\ss e 82, CH-4056 Basel, Switzerland.\\
$^4$Deutsches Elektronen Synchrotron, Notkestra\ss e 85, D-22603 Hamburg, Germany.\\
$^5$Department of Physics, Ain Shams University, Abbassia, 11566 Cairo, Egypt.\\
$^6$Center for Fundamental Physics, Zewail City of Science and Technology, 6 October City, Giza 12588, Egypt.
}
\date{\today}

\begin{abstract}
We investigate the collider signatures of neutral and charged Long-Lived Particles (LLPs), predicted by the Supersymmetric $B-L$ extension of the Standard Model (BLSSM), at the Large Hadron Collider (LHC). The BLSSM is a natural extension of the Minimal Supersymmetric Standard Model (MSSM) that can account for non-vanishing neutrino masses. We show that the lightest right-handed sneutrino can be the Lightest Supersymmetric Particle (LSP), while the Next-to-the LSP (NLSP) is either the lightest left-handed sneutrino or the left-handed stau, which are natural candidates for the LLPs.  We analyze the displaced vertex signature of the neutral LLP (the lightest left-handed sneutrino), and the charged tracks associated with the charged LLP (the left-handed stau).  We show that the production cross sections of our neutral and charged LLPs are relatively large, namely of order ${\cal O}(1)~{\rm fb}$. Thus, probing these particles at the LHC is quite plausible. In addition, we find that the displaced di-lepton associated with the lightest left-handed sneutrino has a large impact parameter that discriminates it from other SM leptons. We also emphasize that the charged track associated with the left-handed stau has a large momentum with slow moving charged tracks, hence it is distinguished from the SM background and therefore it can be accessible at the LHC.  
\end{abstract}
\maketitle

\section{Introduction}
\label{sec1:intro}
Searches for new physics at the LHC are largely based on probing the direct decay of the associated new particles into SM particles, {\it i.e.}, particles are produced and decayed at the interaction point. However, an interesting possibility of revealing new physics is to detect the collider signatures of heavy LLPs. This type of particles has a long lifetime, hence it can travel for a sizable distance inside the detector and decay at a point (secondary vertex) different from its production point (primary vertex). This signal is called displaced vertex, which is one of the essential features of LLPs. These events can be easily observed with almost negligible background. A particle can have a long lifetime, and becomes an LLP, if its couplings with all other particles in its decay channels are extremely small or if the mass difference between this particle and the ones in its decay is very small, so that the phase space is almost closed. Depending on the lifetime of the LLP, its decay may take place either in a detector tracker, or in a calorimeters and muon system, or even outside the detector. 

Various scenarios for the SM extensions, in particularly supersymmetric models, predict the existence of heavy  LLPs with lifetime greater than few nanoseconds and macroscopic decay length. In the MSSM, the LSP is usually the lightest neutralino and the following particles are possible candidates for the NLSP, with quite close mass to the mass of the LSP: Next-to-the lightest neutralino, the lightest chargino, the lightest stau, and the lightest stop. However, all these particles couple to the LSP with either gauge or Yukawa couplings, which are not very suppressed. Also their associated mixing matrices  are not extremely small, so one can easily verify that their decay rates are not very suppressed (larger than $10^{-10}$ GeV), hence their lifetimes are rather short (smaller than nanosecond) unless one assumes a very fine-tuning mass degeneracy between the LSP and the NLSP~{\cite{Kaneko:2011qi,Johansen:2010ac,Kaneko:2008re}. However, it is quite un-natural to have mass degeneracy between particles coming from different sectors in the MSSM, like for instance lightest neutralino and light stau. Therefore, it is tempting to believe that it is quite un-natural to have an LLP in the MSSM. 

The MSSM is the most natural supersymmetric extension of the SM, however, the solid experimental evidence for neutrino oscillations, pointing towards non-vanishing neutrino masses, indicates the departure from the simplest realization of Supersymmetry (SUSY), the MSSM. The BLSSM is a simple extension of the MSSM, based on the gauge group $SU(3)_C \times SU(2)_L \times U(1)_Y \times U(1)_{B-L}$, can account for current experimental results of light neutrino masses and their large mixing~\cite{Khalil:2006yi,Basso:2008iv,Basso:2009gg,Basso:2010yz,Basso:2010as,Majee:2010ar,Li:2010rb,Alves:2015mua,Perez:2009mu,Emam:2007dy,Klasen:2016qux,Khalil:2012gs,Khalil:2013in}.
Within the BLSSM, right-handed neutrino superfields are naturally introduced in order to implement a type I seesaw mechanism, which provides an elegant solution for tiny neutrino masses.  Also, it has been emphasized that the scale of $B-L$ symmetry breaking is related to the SUSY breaking one and both occur in the TeV region~\cite{Khalil:2007dr,Burell:2011wh}. Hence, several testable signals of the BLSSM  are predicted at the LHC~\cite{Abdallah:2016vcn,Abdallah:2015uba}. 

In this paper  we will show that in the framework of the BLSSM, due the decoupling between left-handed and right-handed sectors, if  the LSP is emerged from the right-handed sector and the NLSP (neutral or charged) is coming from the left-handed sector, then the decay of NLSP to the LSP is extremely suppressed and the NLSP becomes naturally an LLP, even if its mass is not degenerate with the LSP. Here, we assume LLP should have decay length larger than 1~mm. In fact, as we will show in the next section, sometimes the mass difference should be larger than or equal to the mass of $W$ (for charged NLSP) or $Z$ (for neutral NLSP) gauge bosons, to open a channel of decay, otherwise the NLSP becomes not only LLP but also an almost stable particle. In particular, we  will show that a notable feature of the BLSSM is that the lightest right-handed sneutrino can be naturally the LSP and stable, so that it is a viable candidate for cold dark matter~\cite{Abdallah:2017gde,Khalil:2011tb,Khalil:2008ps}. We will emphasize that the BLSSM  has a wide range of  parameter space that  allows the probing of the signature of both the lightest left-handed sneutrino and the left-handed stau as neutral and charged LLPs, respectively.  

This paper is organized as follows. In section~2 we discuss the possibility of having one or more LLP in the BLSSM model. In section~3 we provide the basic inputs required for searching for the neutral and charged BLSSM LLPs at the Compact Muon Solenoid (CMS) detector. Section~4 is dedicated to analyzing the signature of the lightest left-handed sneutrino and the left-handed stau, which are our neutral and charged LLPs, respectively. Finally, our conclusions are given in section~5.

\section{Long-lived particles in the BLSSM}
\label{sec2}

The BLSSM is based on the gauge group $SU(3)_C \times SU(2)_L \times U(1)_Y \times U(1)_{B-L}$. 
This model is a natural extension of the MSSM~\cite{Khalil:2007dr} with three
right-handed neutrino superfields ($\hat{N}_i$), to account for measurements of light
neutrino masses, two chiral SM-singlet Higgs superfields ($\hat{\chi}_1$, $\hat{\chi}_2$ with $B-L$
charges $Y_{B-L}=-2$ and $Y_{B-L}=+2$, respectively), whose Vacuum Expectation Values (VEVs)
of their scalar components, $v'_1=\langle \chi_1\rangle$ and $v'_2=\langle \chi_2\rangle$, spontaneously break the $U(1)_{B-L}$, and a vector superfield
necessary to gauge the $U(1)_{B-L}$ ($Z'$) acquires its mass from the
kinetic term of $\chi_{1,2}$: $M^2_{Z'}=g^2_{_{B-L}}v'^2$, where $g_{_{B-L}}$ is the $U(1)_{B-L}$ gauge coupling constant and
 $v'^2 = v'^2_1+v'^2_2$. Therefore, the superpotential of the BLSSM is given by
\bea {W} &=&
(Y_u)_{ij}\hat{Q}_i\hat{H}_2 \hat{U}^c_j+(Y_d)_{ij}\hat{Q}_i\hat{H}_1\hat{D}^c_j+(Y_e)_{ij}\hat{L}_i\hat{H}_1\hat{E}^c_j\nonumber\\
&+& (Y_{\nu})_{ij}\hat{L}_i\hat{H}_2\hat{N}^c_j  + (Y_N)_{ij}\hat{N}^c_i\hat{\chi}_1\hat{N}^c_j\nonumber\\
&+& \mu \ \! \hat{H}_1\hat{H}_2+\mu' \ \! \hat{\chi}_1\hat{\chi}_2.
\label{super-potential-b-l}
\eea%
The $B-L$  charges of the above superfields, the corresponding soft SUSY breaking terms, and the details of $B-L$ radiative symmetry breaking at TeV scale can be found in Ref.~\cite{Khalil:2007dr}.

In this class of models, the sneutrino squared-mass matrix, in the basis  $(\tilde{\nu}_L,\tilde{\nu}_L^\ast,\tilde{\nu}_R,\tilde{\nu}_R^\ast$), is given by~\cite{Elsayed:2012ec}:
\begin{equation}
{\cal M}^2 = \mat{M^2_{LL} ~ &~  M^2_{LR} \\ \\  (M^{2}_{LR})^{\dag} ~ &~  M^2_{RR}
},\label{sneutrino-matrix}
\end{equation}
where
\begin{eqnarray*}
M^2_{LL} \!\! & = &\!\! \left(m_{\tilde{L}}^2+m_D^2+\frac{1}{2}M_Z^2  c_{2\beta}-\frac{1}{2}M_{Z'}^2  c_{2\beta'}\right)\mathbb{I}_2,\\
M^2_{LR} \!\! & = &\!\! m_D(A_\nu-\mu\cot\beta + M_N)~\mathbb{I}_2,\\
(M^2_{RR})_{11}\!\! & = &\!\!(M^2_{RR})_{22}= M_N^2 + m_{\tilde{N}}^2 + m_D^2+ \frac{M_{Z'}^2}{2} c_{2\beta'},\\
(M^2_{RR})_{12}\!\! & = &\!\! (M^2_{RR})_{21}=M_N(A_N-\mu'\cot\beta'),
\end{eqnarray*}
where $c_x\!\!\equiv\!\! \cos x$, $\mathbb{I}_2$ is $2\times 2$ identity matrix, $M_N$ is the right-handed neutrino mass, which is proportional to the $B-L$ symmetry breaking VEV, {\it i.e.}, $M_N= Y_N v'_1 \sim {\cal O}(1)$~TeV, and $m_D=Y_\nu \langle H_2\rangle = Y_\nu v_2$, with $Y_\nu \lsim {\cal O}(10^{-6})$, to fulfill the smallness of  light neutrino masses~\cite{Khalil:2006yi,Basso:2008iv,Basso:2009gg,Basso:2010yz,Basso:2010as,Majee:2010ar,Li:2010rb,Alves:2015mua,Perez:2009mu,Emam:2007dy,Klasen:2016qux,Khalil:2012gs,Khalil:2013in}. The soft SUSY breaking parameters $m_{\tilde{N},\tilde{L}}$ and $A_{\nu,N}$ are the sneutrino, slepton scalar masses and trilinear couplings, respectively, which are given by universal values
at the Grand Unification Theory (GUT) scale and are determined at any scale by their Renormalization Group Equations (RGEs) by
using SARAH~\cite{Staub:2013tta,Staub:2009bi,Staub:2008uz}. Finally $\tan \beta'$ is defined as the ratio between the two $B-L$ VEVs, $\tan\beta'=v'_1/v'_2$, in analogy to the MSSM VEVs $(\tan\beta = v_2 /v_1)$. 

It is worth noting that the mixing between left- and right-handed sneutrinos $(M^2_{LR})$ is quite suppressed, as it is proportional to $Y_\nu \lsim {\cal O}(10^{-6})$, while the mixing between the right-handed sneutrinos and right-handed anti-sneutrinos $(M^2_{RR})$ is quite large, since it is given in terms of $Y_N \sim {\cal O}(1)$. Thus, the eigenvalues of the left-handed sneutrino squared-mass matrix $M^2_{LL}$ are given by
\be 
m^2_{\tilde{\nu}_{L_i}}\!\! = m^2_{\tilde{\nu}^*_{L_i}} \!\! \simeq\ m_{\tilde{L}_{ii}}^2+\frac{M_Z^2}{2} c_{2\beta}-\frac{M_{Z'}^2}{2} c_{2\beta'},~~i=1,2,3. 
\ee
Therefore, depending on the values of the soft scalar masses $m^2_{\tilde{L}_{ii}}$, the corresponding left-handed sneutrino $\tilde{\nu}_{L_i}$ can be light. However, it is important to note that $m^2_{\tilde{L}}$ is constrained by the Lepton Flavor Violation (LFV) limits to be diagonal, which contributes to the charged slepton masses as well~\cite{Hammad:2016bng}. Therefore, 
the lightest left-handed charged slepton $\tilde{\tau}_L$ (the left-handed stau) is almost degenerate with the lightest left-handed sneutrino, $\tilde{\nu}_{L_1}$. Due to the DM constraints, both $\tilde{\tau}_L$ and $\tilde{\nu}_{L_1}$ can not be the LSP. Nevertheless, they can be the NLSP.
\begin{figure*}[t!]
\centering
\includegraphics[width=7.5cm,height=6cm]{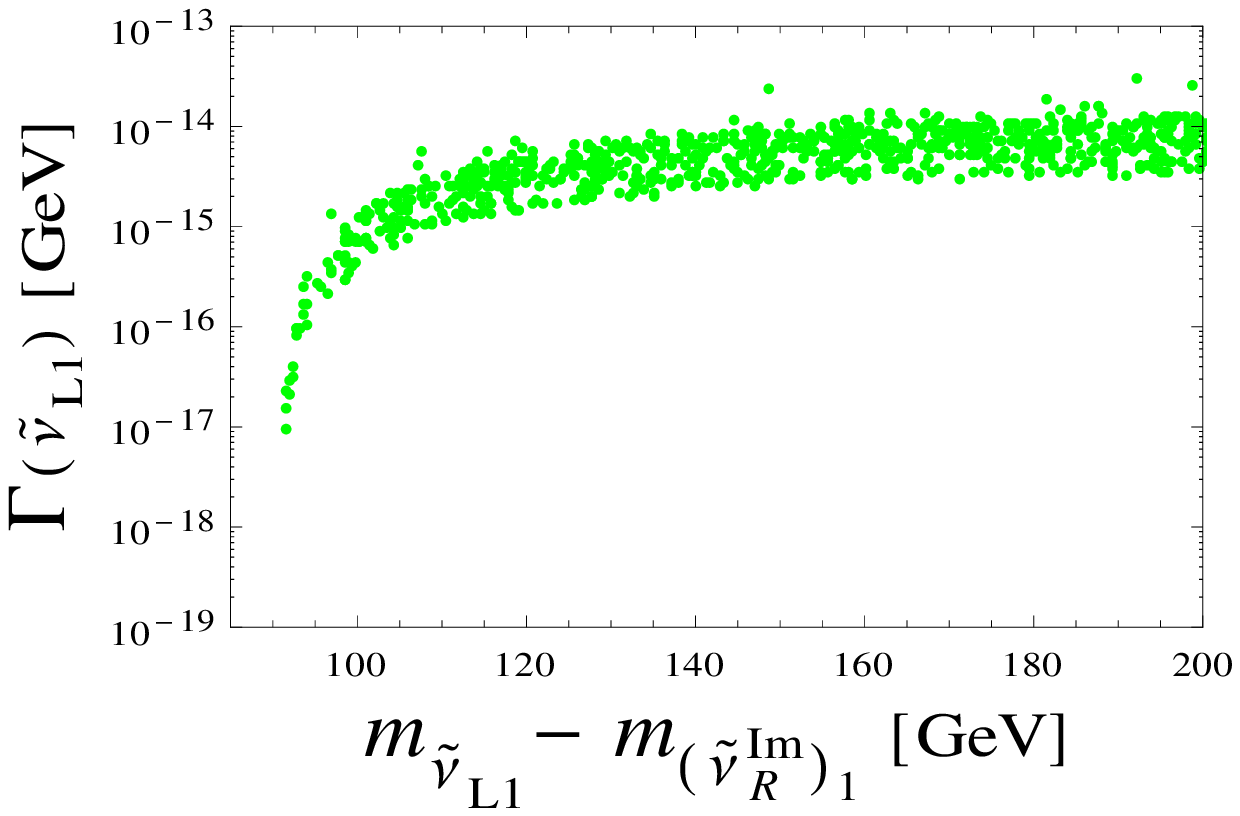}~~~\includegraphics[width=7.5cm,height=6cm]{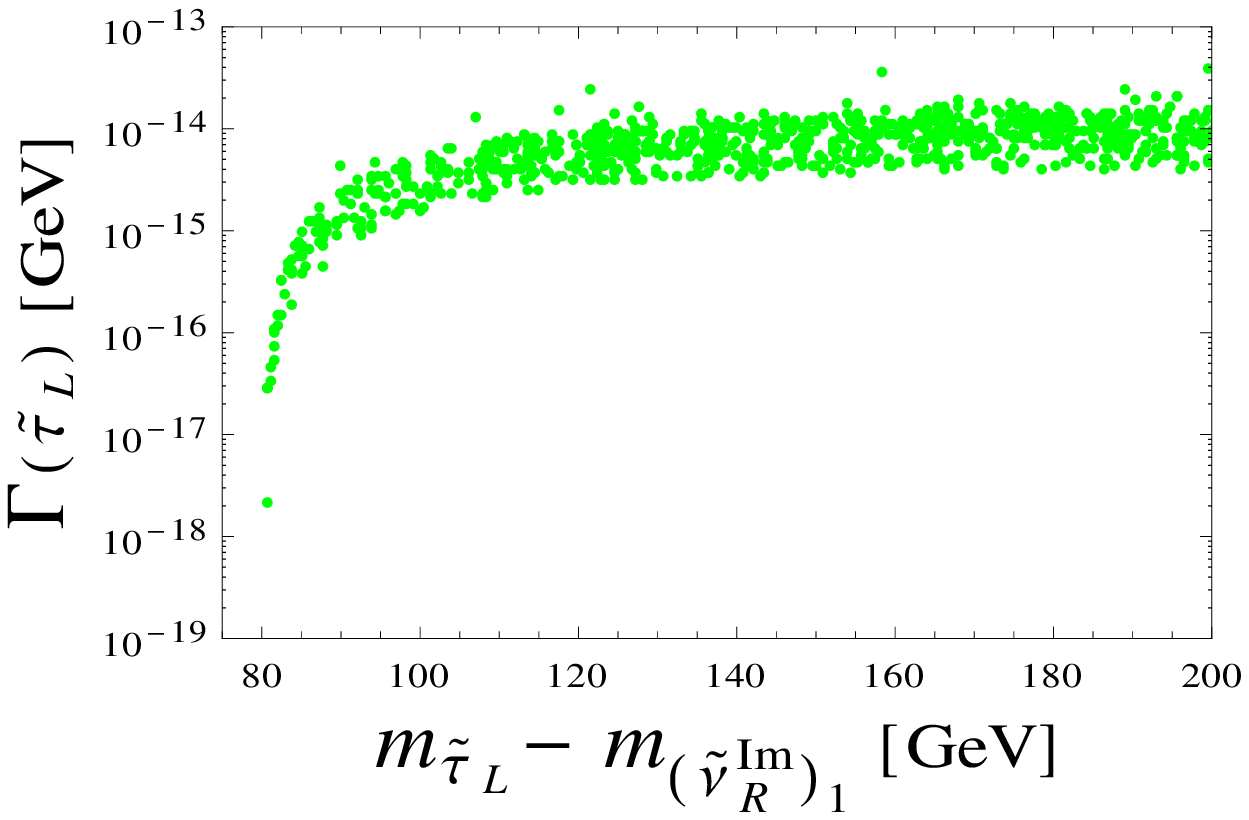}
\caption{Total decay rates of the NLSP including the two- and three-body decays: $\tilde{\nu}_{L_1}$ (left panel) and $\tilde{\tau}_L$ (right panel) as a function of the mass difference $m_{\rm NLSP} - m_{\rm LSP}$, with the LSP is  $(\tilde{\nu}_{R}^{\rm Im})_{_1} $. The points are obtained by the following inputs:
$0.15\le \mu\le 9~{\rm TeV}$, $M_A\simeq 1.8~{\rm TeV}$, $2\le \tan\beta\le 10$, $\mu^\prime\simeq 2.3~{\rm TeV}$, $M_{A^\prime}\simeq 1.4~{\rm TeV}$, $2\le \tan\beta'\le 4$, $g_{BL}\simeq 0.5$, $-0.5\le \tilde{g}\le -0.3$, $M_1\simeq 1.2~{\rm TeV}$, $0.8\le M_2\le 5.5~{\rm TeV}$, $M_3\simeq 1.2~{\rm TeV}$, $0.3\le m_{\tilde{L}_{11(22)}}^2\le 10^3~{\rm TeV}$, $0.1\le m_{\tilde{L}_{33}}^2\le 15~{\rm TeV}$, $ m_{\tilde{N}_{11}}^2\simeq 10~{\rm GeV}$, \ {\rm and} \ $m_{\tilde{N}_{22(33)}}^2\simeq 10~{\rm TeV}$. }
\label{fig:DW}
\end{figure*}
The eigenvalues of the right-handed sneutrino squared-mass matrix $M^2_{RR}$ are given by~\cite{Elsayed:2012ec}:
\begin{equation}
m^2_{\tilde{\nu}_{R}^{\rm Im},\tilde{\nu}_{R}^{\rm Re}} = m^2_{\tilde{\nu}_{R}} \mp \Delta
m^2_{\tilde{\nu}_{R}},
\label{eigenvalues}
\end{equation}
where the mass eigenstates $ \tilde{\nu}_R^{\rm Im}$ and  $\tilde{\nu}_R^{\rm Re}$ are defined as %
\be%
\tilde{\nu}_{R}^{\rm Re} = \frac{1}{\sqrt{2}} \left( \tilde{\nu}_R +
\tilde{\nu}_R^* \right),~~~~~~\tilde{\nu}_{R}^{\rm Im} = \frac{-i}{\sqrt{2}} \left( \tilde{\nu}_R -
\tilde{\nu}_R^* \right).%
\ee %
The squared-mass $m^2_{\tilde{\nu}_{R}}$ is given by%
\be%
m^2_{\tilde{\nu}_{R}}=M_N^2+m_{\tilde{N}}^2+m_D^2+\frac{1}{2}M_{Z'}^2\cos
2\beta',%
\ee%
and  $\Delta m^2_{\tilde{\nu}_{R}}$ is the mass splitting in the
heavy right-handed sneutrinos, which is given by
\be%
\Delta m^2_{\tilde{\nu}_{R}} = M_N \Big\vert A_N-\mu'\cot\beta' \Big\vert. %
\ee%
This mass splitting and the mixing between the right-handed sneutrino
$\tilde{\nu}_R$ and right-handed anti-sneutrino $\tilde{\nu}_R^*$ are proportional to 
$M_N$, which represents the magnitude of lepton number violation. From eq.~(\ref{eigenvalues}), it is clear that the lightest $\tilde{\nu}_{R}$, $(\tilde{\nu}_{R}^{\rm Im})_{_1}$, is typically the lightest sneutrino and can be even the LSP for a wide region of parameter space~\cite{DelleRose:2017ukx}, hence it can be stable and a viable candidate for DM. Depending on the values of $M_N \vert A_N-\mu'\cot\beta' \vert $, the lightest real component of right-handed sneutrino, $(\tilde{\nu}_{R}^{\rm Re})_{_1}$, could be the NLSP.
In this regard, the NLSP can be one of the following particles: $\tilde{\nu}_{L_1}$ (and $\tilde{\tau}_L$), $(\tilde{\nu}_{R}^{\rm Re})_{_1}$ or $(\tilde{\nu}_{R}^{\rm Im})_{_2}$. 
However, among this list of particles only $\tilde{\nu}_{L_1}$ and $\tilde{\tau}_L$ have the chance to be LLPs, while $(\tilde{\nu}_{R}^{\rm Re})_{_1}$ or $(\tilde{\nu}_{R}^{\rm Im})_{_2}$ cannot, as they have quite large decay widths due to the following decay channels: $ (\tilde{\nu}_{R}^{\rm Re})_{_1} \to (\tilde{\nu}_{R}^{\rm Im})_{_1} + Z$ and $ (\tilde{\nu}_{R}^{\rm Im})_{_2} \to (\tilde{\nu}_{R}^{\rm Im})_{_1} + h$. 

In order to investigate explicitly the possibility that $\tilde{\nu}_{L_1}$ and $\tilde{\tau}_L$ can be LLPs, while  $(\tilde{\nu}_{R}^{\rm Re})_{_1}$ or $(\tilde{\nu}_{R}^{\rm Im})_{_2}$ cannot,  we provide here the relevant interaction terms for their dominant decay channels:
\bea
{\cal L}_{\rm int}\!\!&=&\!\!\! \frac{1}{2} \tilde\nu_{L_1} (\tilde\nu_R^{\rm Im})_{_1}(p'\!-p)^\mu Z_\mu (g_1 s_{\theta_W}\!\!+\! g_2 c_{\theta_W})\!\sum_{n=1}^3 R_{1n}I_{1n} \nonumber\\
\!&+&\!\!\! \frac{1}{2} \tilde{\tau}_L \ (\tilde\nu_R^{\rm Im})_{_1} (p'\!-p)^\mu W_\mu~g_2\sum_{n=1}^3 E_{1n}I_{1n}\nonumber\\
\!&+&\!\!\!\frac{1}{2}( \tilde\nu_R^{\rm Re})_{_1} (\tilde\nu_R^{\rm Im})_{_1} (p'\!-p)^\mu Z_\mu~g_{_{B-L}} s_{\theta'}\sum_{n=4}^6 R_{1n}I_{1n}\nonumber \\
\!&+&\!\!\! i ( \tilde\nu_R^{\rm Im})_{_2} (\tilde\nu_R^{\rm Im})_{_1}h~\sum_{k=1}^3\bigg(\frac{1}{4}\Gamma_{13}{T_N}_{kk}-\sqrt{2}\mu'\Gamma_{14}{Y_N}_{kk}\nonumber \\
\!&-&\!\!\!4 v'_1\Gamma_{13}{Y_N}^2_{kk}\bigg) I_{2,k+3}I_{1,k+3},
\label{snu:interactions}
\eea
where $s_x\!\!\equiv\!\! \sin x$, $g_1$ and $g_2$ are the $U(1)_Y$ and $SU(2)_L$ gauge coupling constants, respectively, $\theta_W$ is the Weinberg angle, while $\theta'$ is the mixing angle between the neutral gauge bosons $(Z,Z')$, ${T_N}_{ij}={A_N}_{ij}{Y_N}_{ij}$, and
$\Gamma_{ij}$ and $R_{ij}~(I_{ij})$ are the mixing matrices that diagonalize the CP-even Higgs mass matrix and the CP-even (odd) sneutrino mass matrix, respectively.

It is noticeable that the interactions in the last two terms of eq.~(\ref{snu:interactions}) are proportional to the $B-L$ parameters ($g_{_{B-L}}$ or $Y_N$) and have no significant suppression due to the mixing matrices, since both sneutrinos are from the right-handed sector while the interactions of the first two terms are suffering from stringent suppression due to mis-match between the mixing matrices of left- and right-handed sneutrinos.   

In Fig.~\ref{fig:DW}, we display the total decay rates including the two- and three-body decays, which have been calculated by SPheno~\cite{Porod:2011nf,Porod:2003um}, of the potential BLSSM long-lived particles: $\tilde{\nu}_{L_1}$ (left panel) and $\tilde{\tau}_L$ (right panel) as function of the mass difference $m_{\rm NLSP} - m_{\rm LSP}$, when NLSP is either $\tilde{\nu}_{L_1}$ or $\tilde{\tau}_L$ and the LSP is  $(\tilde{\nu}_{R}^{\rm Im})_{_1} $. Here the points have been generated over the following ranges of the relevant BLSSM fundamental parameters:~$0.15\le \mu\le 9~{\rm TeV}$, $M_A\simeq 1.8~{\rm TeV}$, $2\le \tan\beta\le 10$, $\mu^\prime\simeq 2.3~{\rm TeV}$, $M_{A^\prime}\simeq 1.4~{\rm TeV}$, $2\le \tan\beta'\le 4$, $g_{BL}\simeq 0.5$, $-0.5\le \tilde{g}\le -0.3$, $M_1\simeq 1.2~{\rm TeV}$, $0.8\le M_2\le 5.5~{\rm TeV}$, $M_3\simeq 1.2~{\rm TeV}$, $0.3\le m_{\tilde{L}_{11(22)}}^2\le 10^3~{\rm TeV}$, $0.1\le m_{\tilde{L}_{33}}^2\le 15~{\rm TeV}$, $ m_{\tilde{N}_{11}}^2\simeq 10~{\rm GeV}$, and $m_{\tilde{N}_{22(33)}}^2\simeq 10~{\rm TeV}$. 
As can be seen from this figure, $\tilde{\tau}_L$ and $\tilde{\nu}_{L_1}$ can be long-lived particles only if the difference between their masses and the LSP mass, $m_{(\tilde{\nu}_R^{\rm Im})_{_1}}$, is approximately equal to $M_W$ and $M_Z$,~respectively.
\begin{figure*}[t!]
\centering
\includegraphics[width=6.5cm,height=4.5cm]{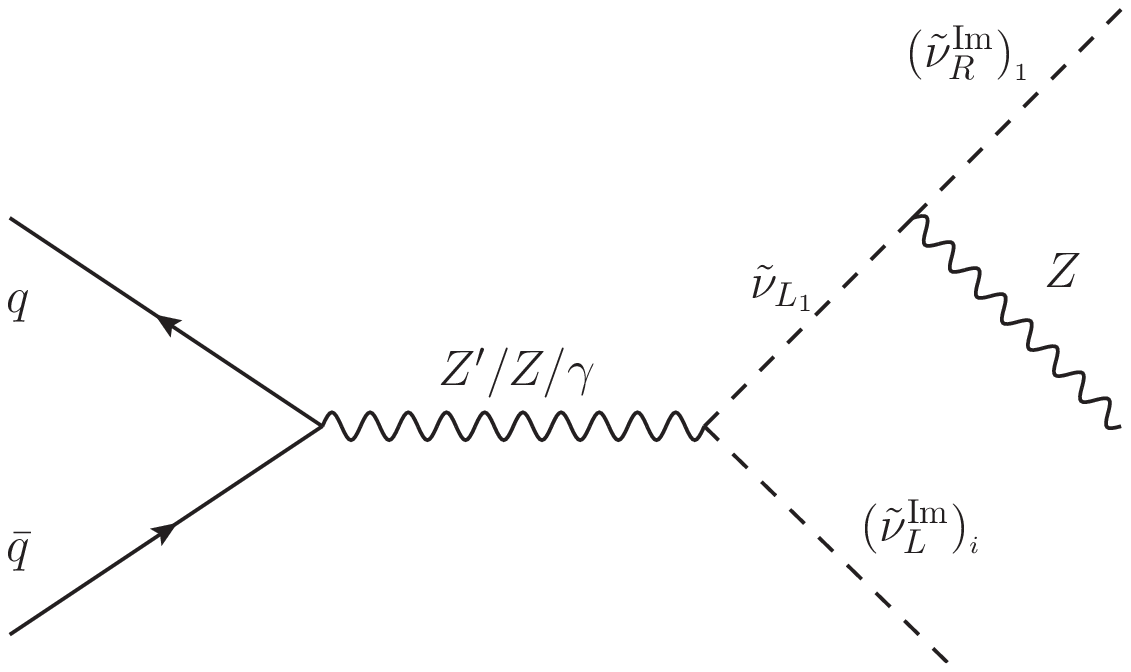}~~\includegraphics[width=6.5cm,height=4.5cm]{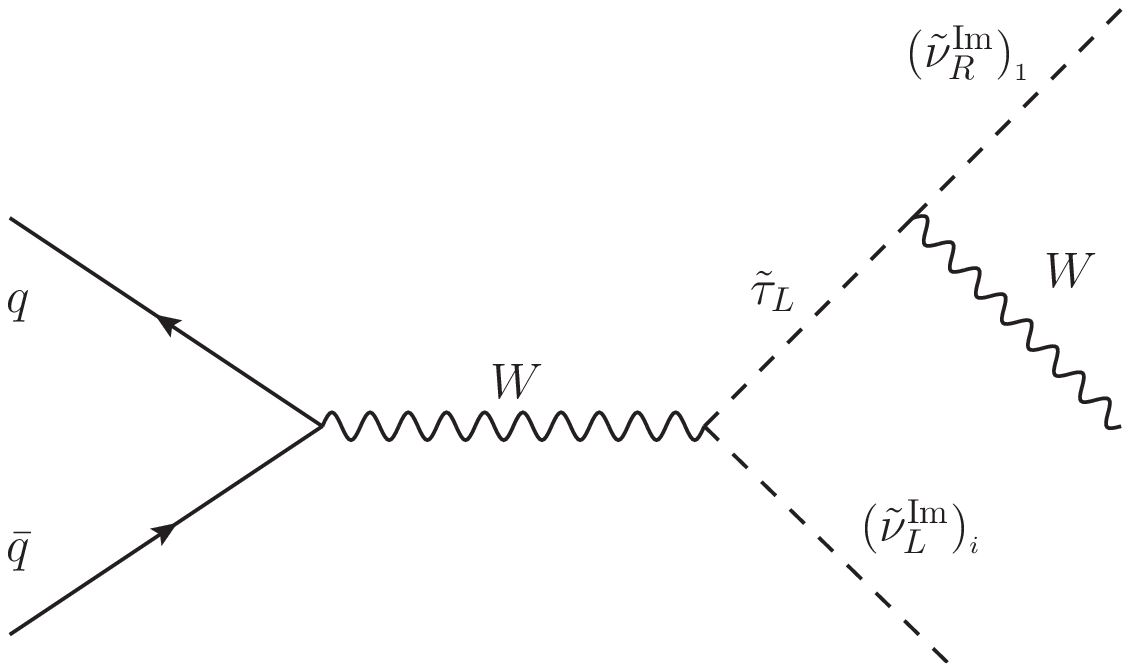}
\caption{Feynman diagrams depicting the production mechanism of the LLPs ($\tilde{\nu}_{L_1}$ and $\tilde{\tau}_L$) at the LHC.}
\label{fyn}
\end{figure*}
\section{BLSSM long-lived particle searches at the CMS}
\label{sec3}

As advocated in the introduction, the main distinct feature of LLPs is that they can travel finite distances before they decay into SM particles. Therefore, these new particles can be probed at the LHC by looking for their displaced vertex signatures. Displaced vertex is defined as the distance between the primary vertex, which is originated from the hard
 scattering at the interaction point, and the secondary vertex that can be identified using the tracker information and the energy loss~\cite{Lenz:2016zpj}. 
In our BLSSM, the lightest left-handed sneutrino  $\tilde{\nu}_{L_1}$, the neutral LLP, can be produced at the LHC through the channel mediated by $Z^\prime/Z/\gamma$ decay, while the left-handed stau $(\tilde{\tau}_L)$, the charged LLP, can be produced through the off-shell $W$ boson, as shown in Fig.~\ref{fyn}. Then later the neutral LLP decays to $(\tilde{\nu}^{\rm Im}_R)_{_1}+Z$, while the charged LLP decays to $(\tilde{\nu}^{\rm Im}_R)_{_1}+W$. We recall that both candidates of our LLPs are from the left-handed sector while the LSP from the right-handed sector, therefore their decay rates are quite suppressed and their lifetime is rather long. 

The efficiency for reconstructing events of displaced vertex depends on the region of the detector where the LLP decays. If the lifetime of the LLP is rather short, then it decays within
the inner detector. For a larger lifetime $\sim {\cal{O}}(1) $ meter, it can decay in the outer layers of the detector, Electromagnetic Calorimeter (ECAL), Hadronic Calorimeter (HCAL) or Muon Chamber (MuC)~\cite{Khachatryan:2015jha,CMS:2014hka}. Finally, if the LLP has a very long lifetime, it may decay outside the detector boundaries. The tracker, which is the innermost detector system of the CMS detector, has a length of $5.8$~m and a diameter of $2.5$~m. It comprises a silicon pixel detector with three barrel layers at radii between $4.4$~cm and $10.2$~cm and a silicon strip tracker with ten barrel detection layers extending outwards to a radius of $1.1$~m~\cite{Chatrchyan:2013sba,Chatrchyan:2014fea}. Each system is completed by endcaps, which consist of two
disks in the pixel detector and three plus nine disks in the strip tracker on each side of the barrel, extending the acceptance of the tracker up to a pseudorapidity of $|\eta|>2.5$~\cite{Chatrchyan:2013sba,Chatrchyan:2014fea}. The observation of a displaced vertex depends on reconstructing the tracks of the charged final state particles. As displayed in Fig.~\ref{displaced vertex}, one needs at least two charged tracks to reconstruct a secondary vertex. Important parameters for this reconstruction are as follows:
\begin{itemize}
\item The total distance from primary to secondary vertices, $L_{xy}$, which can be called the decay length.  This decay length is defined as the distance in $xy$-plane between the particle and its decay products, which is calculated using the generator level information. The decay length $L_{xy}$ is calculated by:
\begin{equation}
L_{xy} = \sqrt{({\rm d}.v(x) - {\rm m}.v(x))^2 + ({\rm d}.v(y) - {\rm m}.v(y))^2 },
\label{eqlxy}
\end{equation}
where mother (m) and  daughter (d) represent the LLP and its decay product, respectively, and $v(x)$ and $v(y)$ represent the vertex coordinates in the $xy$-plane.   
\item The impact parameter, $d_{0} = L_{xy} \sin \theta$, where $\theta$ is the angle described by the trajectory of the displaced vertex with respect to the beam line, which is also depicted in Fig.~\ref{displaced vertex}.
\end{itemize}
\begin{figure}[t!]
\centering
\includegraphics[width=0.35\textwidth]{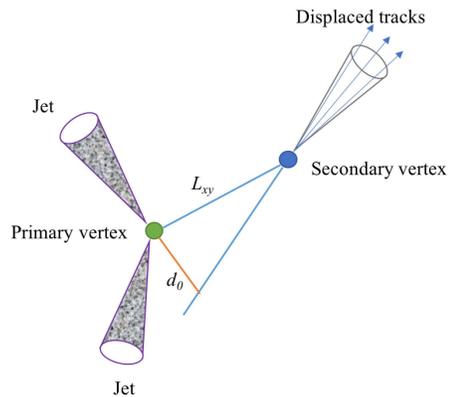}
\caption{A schematic diagram for the LLP displaced vertex.}
\label{displaced vertex}
\end{figure}
In order to distinguish the displaced vertices from primary ones,  the constraints: $|d_{0}| >$~(2--4)~mm and $L_{xy} >$~(4--8)~mm, are usually imposed. With these cuts, the SM backgrounds become almost neglected~\cite{Chatrchyan:2012jna,Aad:2012zx,Cerdeno:2013oya,ATLAS:2012av}. The lifetime of the LLP is defined as 
\begin{equation}
\tau_0 = \frac{\hbar}{\Gamma_{\rm tot.}},
\end{equation}
\begin{figure*}[t!]
\centering
\includegraphics[width=7.5cm,height=6cm]{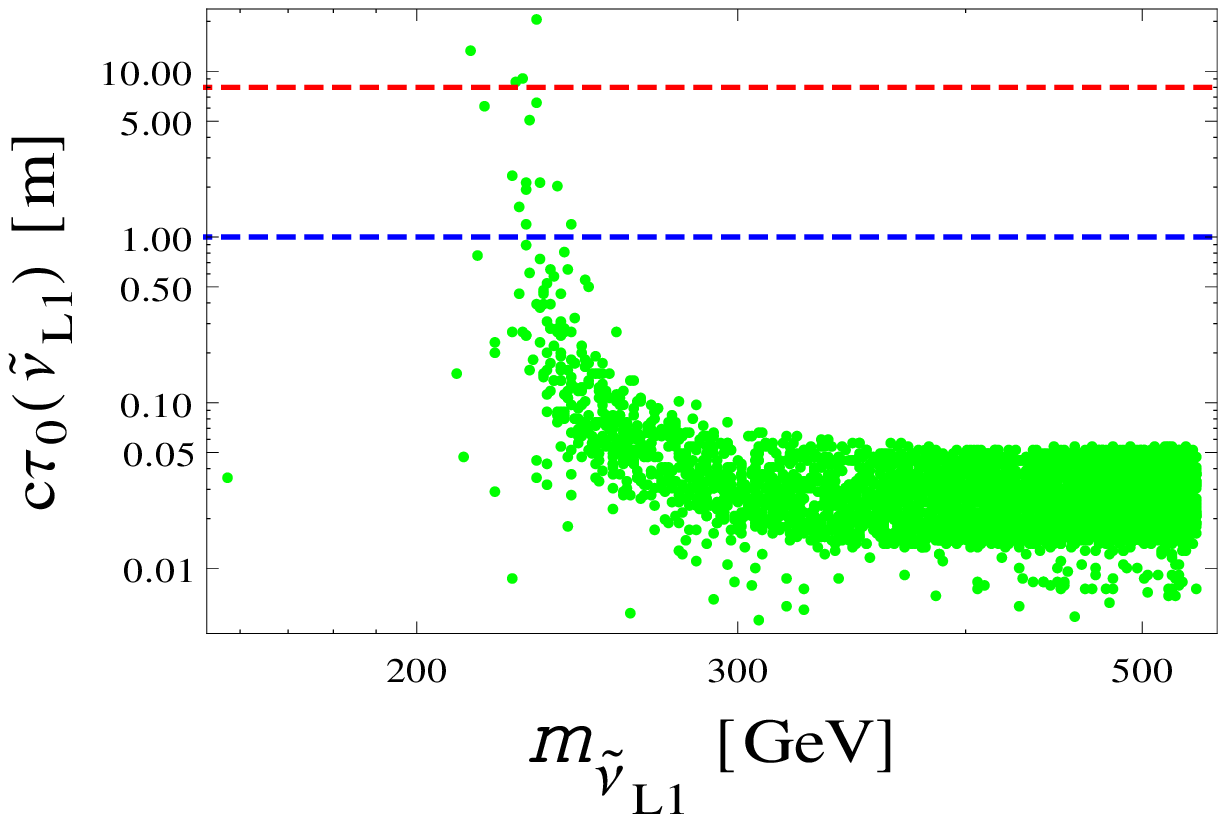}~~~\includegraphics[width=7.5cm,height=6cm]{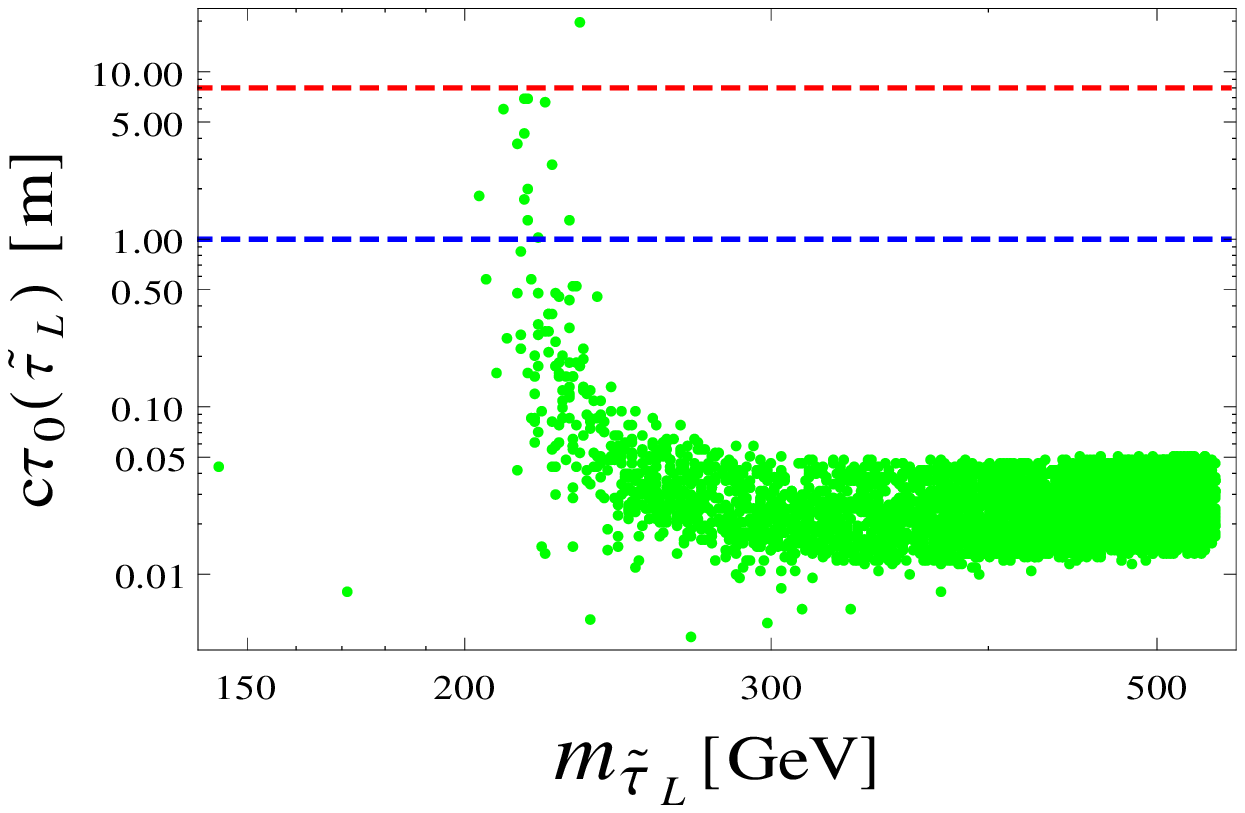}
\caption{Displaced distance of the LLPs: $\tilde{\nu}_{L_1}$ (left panel) and $\tilde{\tau}_L$ (right panel) as function of its mass, where the LSP is  $(\tilde{\nu}_{R}^{\rm Im})_{_1} $. Horizontal (blue and red) lines correspond to the upper limits on the lifetime of the LLP so that particle reaches the ECAL and the MuC. Here, we have used the same inputs of Fig.~\ref{fig:DW}.}
\label{figlifetime}
\end{figure*}
where $\Gamma_{\rm tot.}$ is the total decay width of the LLP.  As mentioned, the total decay width is proportional to the mixing between both left- and right-handed sectors which is highly suppressed. Therefore, the lightest left-handed sneutrino and the left-handed stau will have large lifetimes. Fig.~\ref{figlifetime} shows the displaced distance of the neutral LLP (left panel) and the charged LLP (right panel) versus the mass of the corresponding} LLP. It is clear that in both cases, the neutral and charged LLPs, one can get a significant traveled distance by the LLP before its decay within the detector boundaries, where, in Fig.~\ref{figlifetime}, the blue and red dashed lines correspond to the dimensions of the ECAL and the MuC, respectively. It is worth noting that we have used combined information from the tracker, the ECAL and the MuC to study leptons from LLP decays.

The LLP lifetime $\tau$ in the laboratory frame is related to the proper
lifetime $\tau_0$ as
\begin{equation}
\tau = \gamma \tau_0,
\end{equation}
where $\gamma$ is the Lorentz factor, which depends on mass and momentum of the LLP.
Depending on its lifetime, the LLP can travel a finite distance after its production at the LHC and before its decay.

\begin{figure*}[t!]
\centering
\includegraphics[width=7.5cm,height=6cm]{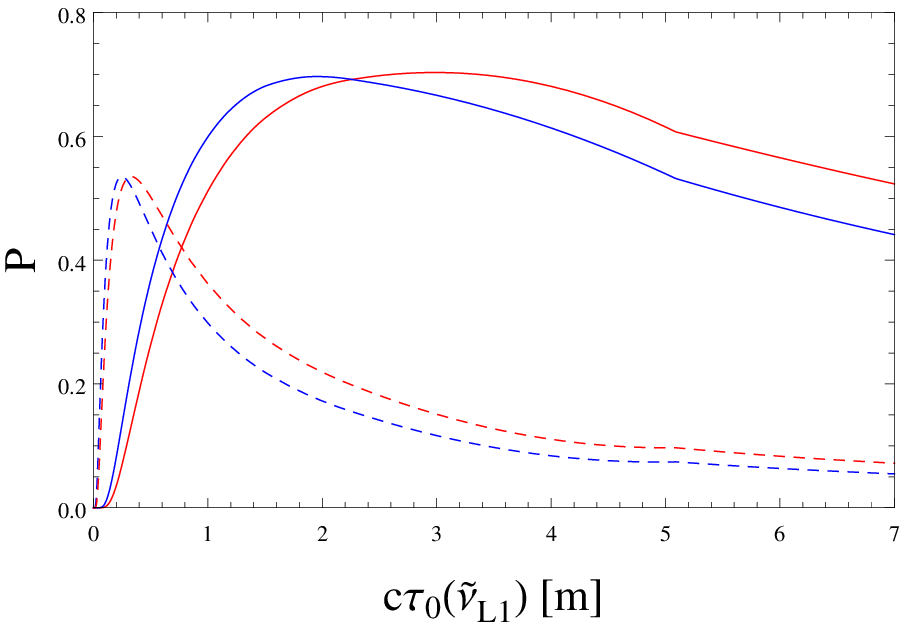}~~~\includegraphics[width=7.5cm,height=6cm]{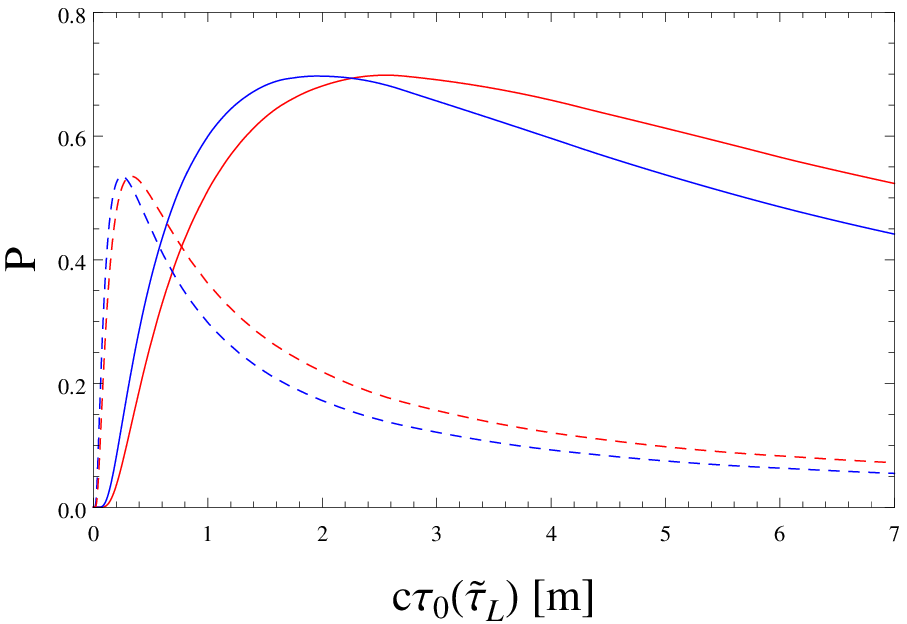}
\caption{Probability distribution of the neutral LLP (left panel) and the charged LLP (right panel) to decay within the detector regions~$a$ and $b$, as explained in the text, versus their displaced distance $c\tau_0$. Here, red and blue curves denote $\gamma=0.75$ and $\gamma=1$, respectively, and solid curves refer to region~$a$ while dashed curves refer to region~$b$. Again, we have used the same inputs as in Fig.~\ref{fig:DW}.}
\label{fig1}
\end{figure*}

Besides the traveled distance, the probability of the long-lived particles decaying in the detector depends on their kinematic configuration and on the region of
the detector in which the decay may occur. This probability can be defined in a region determined by inner and outer distances of the detector, where the LLPs decay to their SM components, as follows~\cite{Accomando:2016rpc,Antusch:2016vyf}
\begin{equation}
P = \int_{r_1}^{r_2} dx \frac{1}{c\tau}  e^{(-\frac{x}{c\tau})},
\label{eq}
\end{equation}
where $r_1$ and $r_2$ are both inner and outer distances of the detector, which depend on pseudorapidity $\eta$, and $c$ is the speed of light. Fig.~\ref{fig1} shows probability distribution for the decay of the neutral LLP (left panel) and the charged LLP (right panel) in two regions ($a$ and $b$): Region~$a$, where $r_1=0.5$~m and $r_2=5$~m, has been chosen to select LLPs that decay in the tracker beyond where tracks can be reconstructed and before the MuC to provide the possibility of observing muon hits. Region~$b$, where $r_1=0.1$~m and $r_2=0.5$~m, has been selected to probe LLPs that decay within the inner tracker. Also, one can see that, at high $c\tau_0$ the probability of finding a muon pair in region~$a$ is dominated (solid curves), while for low $c\tau_0$ the probability for finding electron pair in region~$b$ can exceed (dashed curves). Thus, in our analysis, to reconstruct the LLP we have required the existence of both electron and muon pairs. The expected number of events for long-lived decays after it moves a distance in the detector between $r_1$ and $r_2$ can be obtained by
\begin{equation}
N = P \ {\cal{L}} \ \sigma_{\rm tot.},
\end{equation}
where $P$ is the probability of finding a long-lived particle decay and is defined in eq.~(\ref{eq}), ${\cal{L}}$ is the LHC integrated luminosity and $\sigma_{\rm tot.}$ is the total production cross section for the full process at the LHC.
\section{BLSSM long-lived signatures at the CMS}
\label{sec4}
In this section we discuss the signature of the above mentioned examples of neutral and charged LLPs ($\tilde{\nu}_{L_1}$ and $\tilde{\tau}_L$) by using simulated events in proton-proton collisions at $\sqrt{s} = 13$~TeV and with integrated luminosity 100~fb$^{-1}$. The matrix-element calculation and the generation of signal events at parton level
have been done using {\tt MadGraph5\_aMC@NLO}~\cite{Alwall:2014hca}, whereas, for parton showering as well as hadronization, we have used {\tt PYTHIA8}~\cite{Sjostrand:2007gs}. The detector simulation has been performed using {\tt DELPHES-v3.4.1} package~\cite{deFavereau:2013fsa}. Finally, the analysis and plots have been produced using {\tt ROOT6} object orientated data analysis framework~\cite{Brun:1997pa}. Every time the CMS proton bunches cross one another, more than one proton-proton collision takes place; this is known as pile-up. In order to take the pile-up effects into account, all samples have been simulated with pile-up = 55 per vertex.

The neutral LLP, $\tilde{\nu}_{L_1}$, is characterized by the $Z$ boson decays to displaced di-lepton in the final state, while the charged LLP, $\tilde{\tau}_L,$ is characterized by the charged tracks properties which are different in energy deposit in the tracker and in the momentum from those associated with the SM processes. The highly displaced tracks can be reconstructed in the CMS tracker~\cite{Chatrchyan:2014fea}. This ability allows  the study of the neutral LLP tracks, where it cannot be probed using its tracks information itself, but it can only be probed using its displaced tracks information formed by the charged decay products of this LLP.  At the moment, the CMS is less sensitive to probe the LLPs with decay length 
$c\tau_0 > 1$~m~\cite{CMS:2014hka}, thus for the neutral LLP $\tilde{\nu}_{L_1}$, we consider two cases: $c\tau_0 = 6.5$~m and $c\tau_0 = 90$~cm, with production cross sections: $7.6$~fb and $4.9$~fb, respectively, while for the charged LLP $\tilde{\tau}_L$, we also consider two cases: $c\tau_0 = 6.9$~m and $c\tau_0 = 85$~cm, with production cross sections: 9.6~fb and 1.7~fb, respectively. This set of points has been chosen from the random scan in Fig.~\ref{fig:DW} to analyze the LLPs decaying in two different regions of the CMS detector, outside the tracker ($c\tau_0>1$~m) and inside the tracker ($c\tau_0\lesssim 1$~m).
\vspace{-0.5cm}
\subsection{Long-lived sneutrino signature at the CMS}
\begin{figure}[t!]
\centering
\includegraphics[width=7.5cm,height=5.5cm]{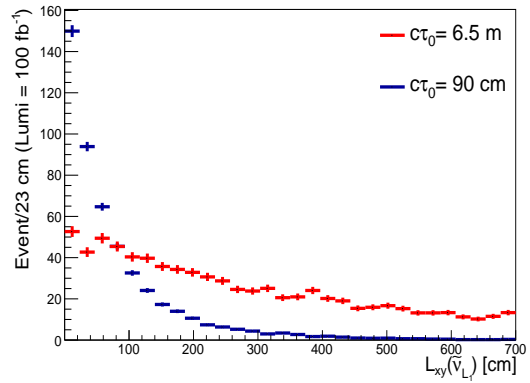}
\caption{$L_{xy}$ for the neutral LLP, $\tilde{\nu}_{L_1}$, which is calculated from eq.~(\ref{eqlxy}).}
\label{LLPGen}
\end{figure}

We start with analyzing the signal of the neutral LLP, $\tilde{\nu}_{L_1}$, which decays to $Z$ boson and $(\tilde{\nu}^{\rm Im}_R)_{_1}$ as a missing transverse energy ($\met$), as shown in Fig.~\ref{fyn}  (left panel). The experimental signature is a distinctive topology consisting of a pair of charged leptons originating from a displaced secondary vertex~\cite{CMS:2014mca}. Due to the fact that all the SM particles should have a decay length $L_{xy}$ of order $4$~mm, we require $L_{xy}> 20$~mm  in order to distinguish between our signal and the SM background. In Fig.~\ref{LLPGen}, we show $L_{xy}$ distribution for the neutral LLP, $\tilde{\nu}_{L_1}$. As can be seen from this figure, our neutral LLP can travel up to $7$~m in the $xy$-plane of the CMS detector.   

\begin{figure}[t!]
\centering
\includegraphics[width=7.5cm,height=5.5cm]{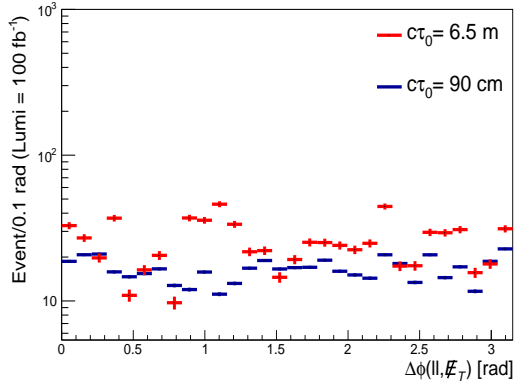}
\caption{The angular separation in $\phi$ between di-lepton and $\met$.}
\label{Dphillmet}
\vspace{-0.5cm}
\end{figure}

As emphasized, the SM particles could have a  displaced di-lepton with much smaller $d_0$ than those of the LLP~\cite{CMS:2016isf}. Therefore, the event selection focuses exclusively on a displaced and isolated lepton signature and it does not try to identify signal events using missing energy. In this way, we retain sensitivity to our model which can produce leptons with displacements from $100~\mu$m to more than $20$~cm, regardless whether these leptons are accompanied  $\met$, or other interesting features. In Fig.~\ref{Dphillmet}, we show the angular separation in the $\phi$ direction between $ll$ and $\met$, $\Delta\phi(ll,\met)$, where $l=e,\mu$. Since the signal has two sources of $\met$ as shown in Fig.~\ref{fyn} (left panel), $\Delta\phi$ is almost flat distribution and we do not apply any cut on it in our selection. However, such a cut may play a role in separation between our signal and any SM process could have a $\met$ in a specific direction.

Moreover, in our analysis, if isolated leptons (muons or electrons) are matched to a track within $\Delta R = \sqrt{(\Delta\eta)^2+(\Delta\phi)^2}$ < 0.1, then this lepton track is obtained, where $\Delta\eta$ and $\Delta\phi$ represent the angular separation between leptons and tracks in $\eta$ and $\phi$ coordinates, respectively. In order to separate our signal from SM contamination we apply additional cuts on leptons as follows: 
\begin{enumerate}
\item  In the di-muon channel: two oppositely charged muons with $p_T(\mu)$ > 26~GeV.
\item In the di-electron channel: two  charged electrons with $p_T(e)$ > 36~GeV while charge requirement in case of electrons is relaxed, as bremsstrahlung can result in incorrect charge reconstruction especially of high $p_T$ electrons~\cite{CMS:2014mca}. 
\end{enumerate} 

\begin{figure}[t!]
\centering
\includegraphics[width=8cm,height=6cm]{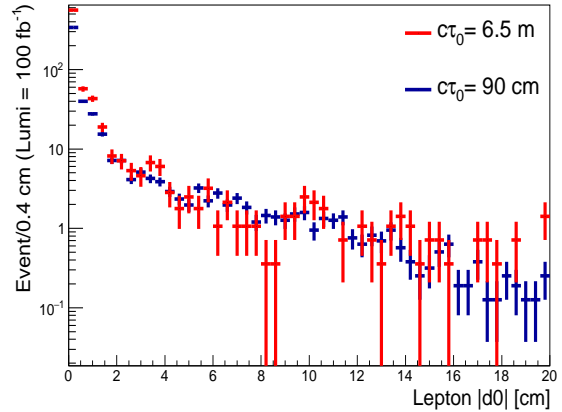}
\caption{Impact parameter distribution for the displaced electron or muon pair produced from $Z$ decay at the ECAL or the MuC, respectively.}
\label{nuetrald0}
\end{figure}

After applying all these selections, the neutral LLP signature is characterized by two isolated and displaced leptons in the final state. These two leptons should have a large impact parameter $d_0$ which separates them from the leptons produced from SM processes. The number of signal events after each cut shown in table~\ref{cut1}. It is worth mentioning that, with this set of cuts, our signal is background free.  In Fig.~\ref{nuetrald0}, we show the number of events of the signal versus the impact parameter $d_0$, which can be larger than 10~cm for leptons. The region with small $d_0$, {\it i.e.}, $d_0 \simeq$ few hundreds of $\mu$m  is expected to be contaminated by the SM background~\cite{CMS:2014hka,Lara:2018rwv}, while at the region with larger $d_0 \simeq$ few cm is expected to be dominated by our signal for the neutral LLP, $\tilde{\nu}_{L_1}$.  
\begin{table}
\begin{center}
\begin{tabular}{|c|c|c|}
\hline 
Cuts& $c\tau_0 = 90$~cm & $c\tau_0 = 6.5$~m \\ 
\hline
Before cuts& $4960000$ & $7600000$ \\
\hline 
$n(l)\geq 2$ & $561463$ & $898899$ \\ 
\hline 
$p_T(\mu)>26$~GeV $\&$ $p_T(e)>36$~GeV &  $176828$ & $86971$ \\ 
\hline 
$\Delta R(l,{\rm track}) < 0.1 $ & $238$ & $114$ \\  
\hline 
$d_0 > 300~\mu m$ & $223$ & $107$ \\
\hline
\end{tabular}
\caption{Signal cut-flow for the two cases of the neutral LLP: $c\tau_0 = 90$~cm and $c\tau_0 = 6.5$~m, normalized to its total cross section and the integrated luminosity.}
\label{cut1}
\end{center}
\end{table}
\begin{figure}[t!]
\centering
\includegraphics[width=7.5cm,height=6cm]{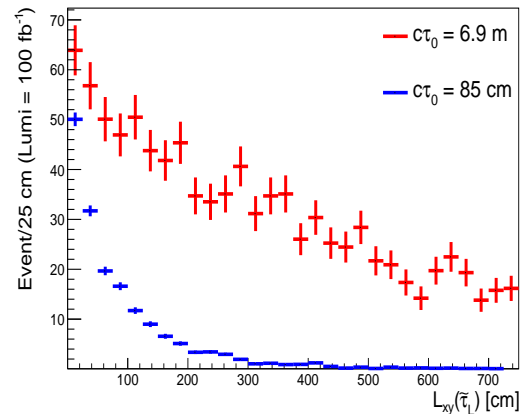}
\caption{$L_{xy}$ for the charged LLP, $\tilde{\tau}_L$, which is calculated from eq.~(\ref{eqlxy}).}
\label{LLCP1}
\end{figure}
\subsection{Long-lived stau signature at the CMS}
\label{sec5}
We now turn to the charged LLP, $\tilde{\tau}_L$, analysis. While the displaced di-lepton gives a clear signature for the neutral LLP, the track analysis is used for the charged LLP signature, which behaves like heavy and slow moving muon. The velocity of the charged LLP, is defined as~\cite{Ghosh:2017vhe}
\begin{equation}
\beta = \frac{p}{E},
\end{equation}
where $E$, $p$ are energy and momentum of the track associated with the charged LLP and are  considerably lower than unity, which is not the case for muons. Another characteristic feature of $\tilde{\tau}_L$ as the charged LLP is that its associated tracks have large momenta, which can be used to separate them from the muons. In this study, motivated by the analysis in Ref.~\cite{Khachatryan:2016sfv}, cuts are performed on transverse momentum $p_T$, and $\beta$ for heavy charged tracks to distinguish them from muons. They are as follows:
\begin{figure*}[t!]
\centering
\includegraphics[width=8cm,height=6cm]{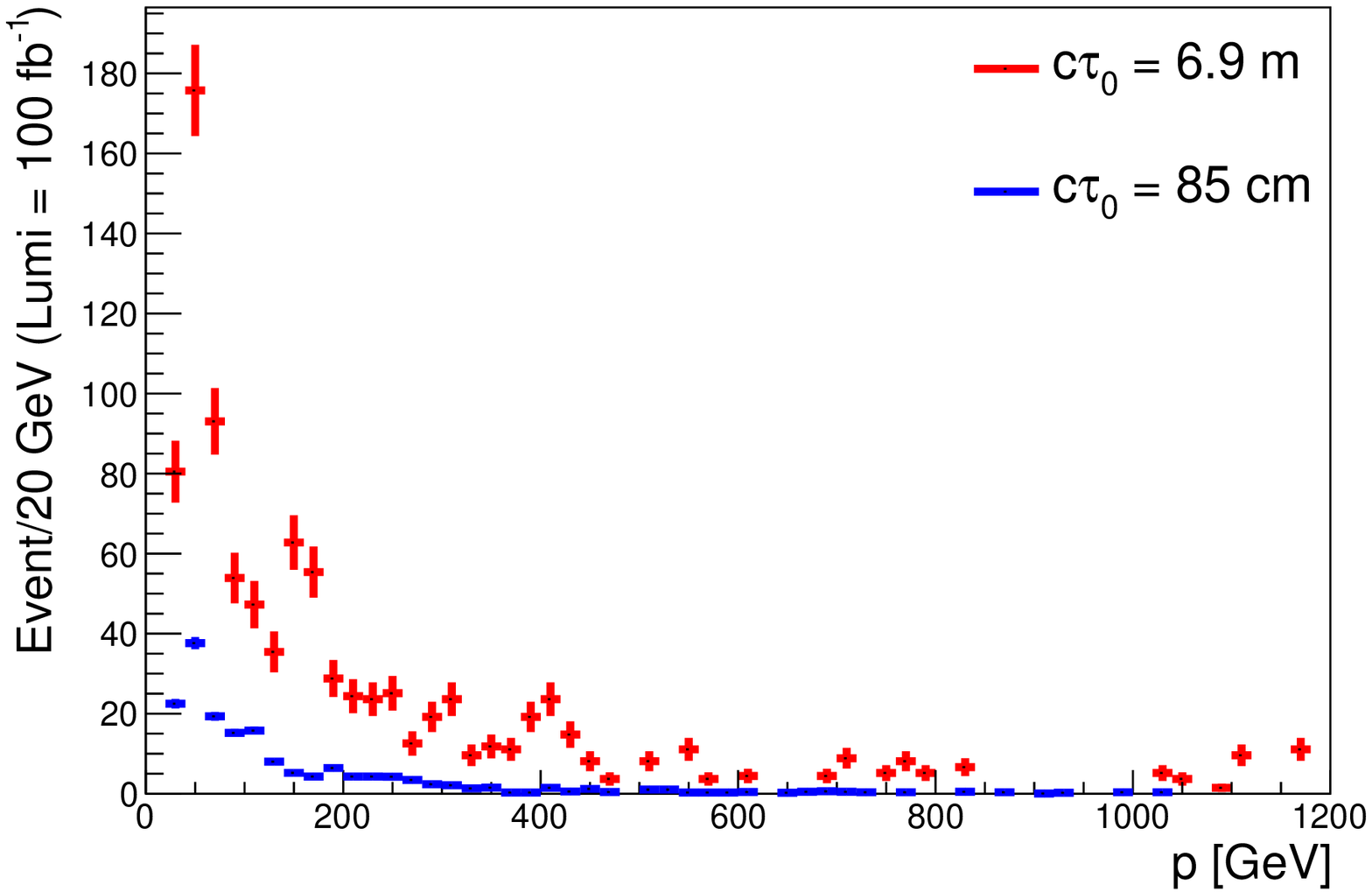}~~\includegraphics[width=8cm,height=6cm]{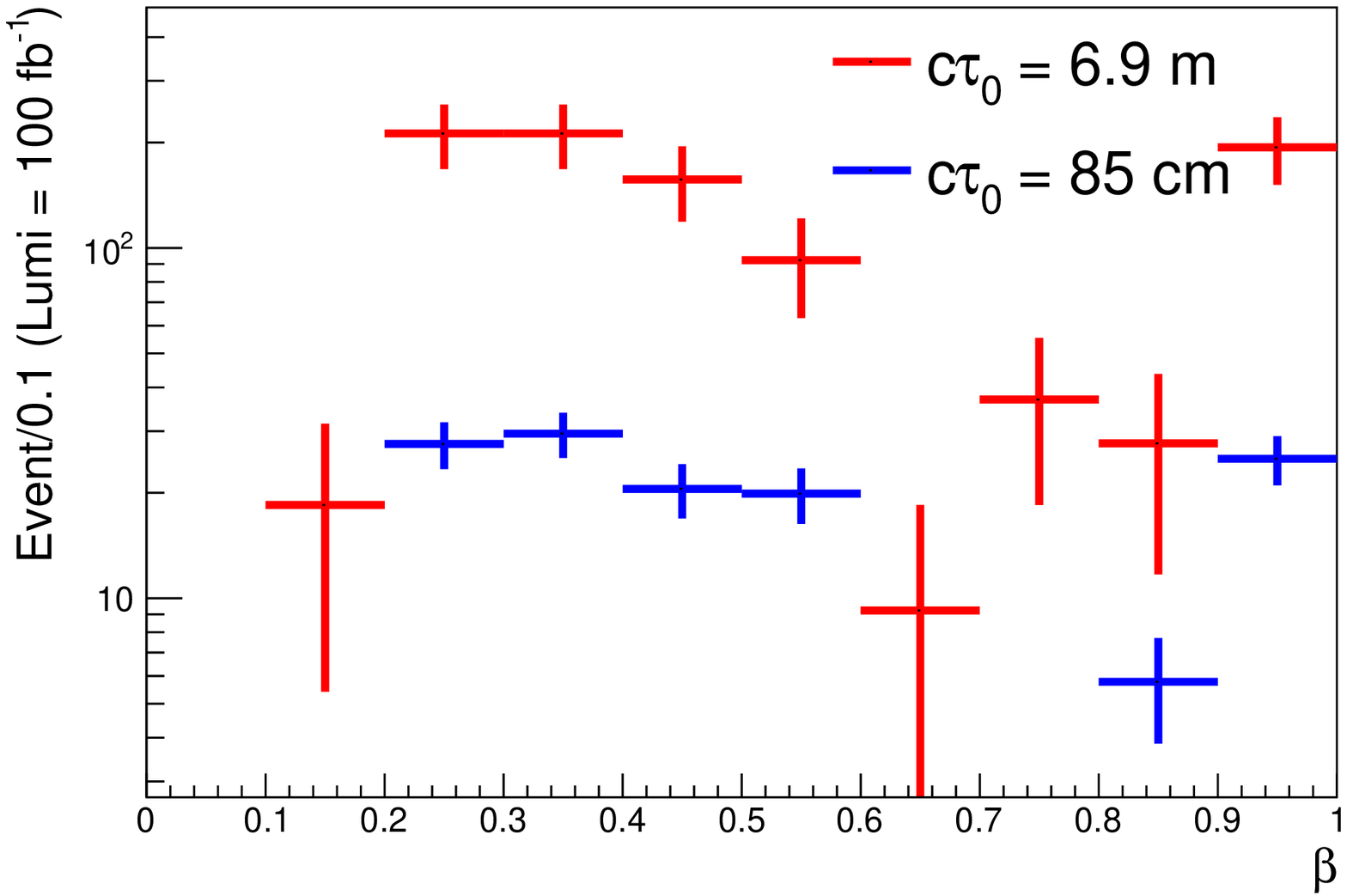}
\caption{Left panel, the momentum $p$ of the track associated with the charged LLP, $\tilde{\tau}_L$. Right panel, $\beta$ distribution for the charged LLP, which is calculated from the track information.}
\label{LLCP2}
\end{figure*}
\begin{enumerate}
\item As for the neutral LLP case, $L_{xy}$ is used to reject some of the SM backgrounds which have $L_{xy}$ $\lesssim$ 4~mm. Here also, we select $L_{xy} > 20$~mm as shown in Fig.~\ref{LLCP1} which indicates a large $L_{xy}$ (up to 7~m) for our charged LLP. 
\item The tracks associated with the charged LLP are chosen with $p_T > 30$~GeV and $|\eta| < 2.1$. This is consistent with the relatively high track $p$ of our $\tilde{\tau}_L$, as shown in Fig.~\ref{LLCP2} (left panel). As emphasized, this is one of the characteristics of the charged~LLP. 
\item An extra cut: $\Delta R(\tilde{\tau}_L,{\rm track})$ < 0.5 is imposed to confirm that a specific track originated from the charged LLP $\tilde{\tau}_L$ not from other sources. This cut is applied using the information from tracks and generator level variables with $\tilde{\tau}_L$ selected with its pdgId~\cite{Patrignani:2016xqp}.
\item We impose the cut $\beta < 0.95$ to distinguish our charged LLP from muons. Small values of $\beta$ for $\tilde{\tau}_L$ are quite natural as shown in Fig.~\ref{LLCP2} (right panel), where there is a significant number of events for such small $\beta$.
\end{enumerate}

After applying all of the above cuts except $\beta < 0.95$, the remaining tracks associated with the charged LLP have a large track $p$. After applying the $\beta$ cut, the remaining tracks have a small $\beta$ compared with the muons. This small $\beta$ means that the ratio between the traveling velocity of the charged LLP and the speed of light is significantly smaller than unity, as expected. Table~\ref{cut2} shows the signal cut flow for the two benchmark points. It is worth mentioning that, with this set of cuts our signal is background free. Using those two variables, track $p$ and $\beta$, we can discriminate the charged LLP from other SM backgrounds which are expected to have relatively small track $p$ and $\beta\simeq 1$.

\begin{table}
\begin{center}
\begin{tabular}{|c|c|c|}
\hline 
Cut& $c\tau_0 = 85$~cm & $c\tau_0 = 6.9$~m \\ 
\hline
Before cuts & $1700000$ & $9600000$ \\
\hline 
$L_{xy} > 20$~mm & $3701$ & $20782$ \\ 
\hline 
Track ${p_T}(\tilde{\tau}_L) > 30$ GeV &  $2062$ & $11793$ \\ 
\hline 
$\Delta R(\tilde{\tau}_L,{\rm track}) < 0.5 $ & $52$ & $745$ \\  
\hline 
$\beta < 0.95 $& $50$ & $717$ \\
\hline
\end{tabular}
\caption{Signal cut-flow for the two cases of the charged LLP: $c\tau_0 = 85$~cm and $c\tau_0 = 6.9$~m, normalized to its total cross section and the integrated luminosity.}
\label{cut2}
\end{center}  
\end{table}
\section{Conclusions}

In this paper, we studied two examples of possible LLPs, predicted by the BLSSM, namely the lightest left-handed sneutrino, $\tilde{\nu}_{L_1}$, and the left-handed stau, $\tilde{\tau}_L$, which act as neutral and charged LLPs, respectively. A salient feature of these NLSP particles is that they do not need to be degenerate in mass with the LSP. Thus, a significant fine-tuning, usually assumed in the MSSM and other LLP scenarios, is avoided. Their long lifetimes are mainly due to the fact that they emerged from the left-handed sector, while the LSP of the considered model emerged from the  right-handed sector, therefore their couplings with the LSP are extremely suppressed. 
The main production of these LLPs at the LHC is through the processes: $p p \to Z'/Z/\gamma \to \tilde{\nu}_{L_1} + (\tilde{\nu}^{\rm Im}_{R})_1$ and $p p \to W \to \tilde{\tau}_L + (\tilde{\nu}^{\rm Im}_{R})_1$.  Then, they decay into $\tilde{\nu}_{L_1} \to (\tilde{\nu}^{\rm Im}_{R})_1 + Z$ and $\tilde{\tau}_L \to (\tilde{\nu}^{\rm Im}_{R})_1 +W$. To make the lifetime of these particles not very long, so that they can decay inside the detector, the difference between their masses and the LSP mass should be larger than or equal to $M_Z$ (for the neutral LLP) and $M_W$ (for the charged LLP).  

We have analyzed the signatures of these LLPs at the LHC, for a run with a center of mass energy $\sqrt{s}=13$ TeV and integrated luminosity = $100~{ \rm fb}^{-1}$. The impact parameter $d_0$ which is the main characteristic variable for displaced vertices analysis has been reconstructed for the neutral LLP signal using a fast detector simulator. We have shown that the total decay width for  $\tilde{\nu}_{L_1}$  can reach $10^{-17}$ GeV, which enables it to have a large $d_0$, $20$~cm or even more, hence the associated signal can be easily identified from the SM backgrounds. Also the total decay width of the $\tilde{\tau}_{L}$ can reach  $10^{-17}$ GeV, so that the analysis for  the charged LLP $\tilde{\tau}_{L}$ carried out with cuts on the associated tracks are imposed to choose high $p$ and slow moving charged tracks.

After combining all results, we conclude that, unlike MSSM the $B-L$ extension of MSSM naturally provides long-lived candidates without any fine-tuning, with clean signatures that can be reachable at the next run of the LHC. 

\label{sec4:conc}
\section*{Acknowledgements}
The authors would like to thank Ahmed Abdelalim and Nicola De Filippis for fruitful discussions. A.K. would like to thank Akshansh Singh and Isabell Melzer-Pellmann for useful discussions about the long-lived analysis. The work of W.A. and S.K. is partially supported by the STDF project 18448 and the European Union FP7 ITN INVISIBLES (Marie Curie Actions, PITN-GA-2011-289442). W.A. would like to thank the Department of Atomic Energy (DAE) Neutrino Project of Harish-Chandra Research Institute. The work of A.H. is supported by the Swiss National Science Foundation and the work of A.K. is supported by the Helmholtz Association of German Research Centres. 


\end{document}